\documentclass[twocolumn,twoside]{svmultivs_gm} 
\usepackage{graphicx}	
\usepackage{amsmath}	
\usepackage{multicol}        
\usepackage{bm}		
\usepackage{pdflscape}	
\usepackage[T1]{fontenc}
\usepackage{ae,aecompl}
\usepackage{newtxtext,newtxmath}
\usepackage{blindtext}
\usepackage[whole]{bxcjkjatype}
\usepackage[normalem]{ulem}
\usepackage{lineno}
\usepackage{here}
\usepackage{color} 
\usepackage{multirow}
\usepackage{ulem}
\usepackage{float}
\usepackage{setspace}
\usepackage{booktabs} 
\usepackage{xcolor}

\newcommand{\SGR}{Sgr~A$^\ast$~}
\newcommand{\virgoa}{M\,87\hspace{0.3mm}}
\renewcommand{\textbf}[1]{{#1}}
\newcommand{\jm}[1]{{}}
\newcommand{\jmold}[1]{{}}

\newcommand{\DEG}{^{\circ}}

\title*{Black hole ring images from PSF structures}
\titlerunning{Black hole images from PSF}
\author{Makoto Miyoshi$^1$, Yoshiaki Kato$^2$, \& Junichiro Makino$^3$
}
\authorrunning{Last Name(s) of Author(s)} 
\authoremails{
makoto.miyoshi@nao.ac.jp,
yoshi\_kato@met.kishou.go.jp,
makino@mail.jmlab.jp
}
\institute{1. National Astronomical Observatory Japan, Tokyo, Japan,
\\ 
2. Japan Meteorological Agency, Tokyo, Japan,\\
3. Kobe University, Hyogo, Japan
}
\ContactAuthorName{Makoto Miysohi}
\ContactAuthorTelephone{+81-422-34-3937}
\ContactAuthorEmail{makoto.miyoshi@nao.ac.jp/makoto.miyoshi@icloud.com}
\NumberofInstitutions{3}
\InstitutionPostAddress{1}{N.A}
\InstitutionCountry{1}{N.A}
\InstitutionWebPage{1}{N.A}
\InstitutionPostAddress{2}{N.A}
\InstitutionCountry{2}{N.A}
\InstitutionWebPage{2}{N.A}
\begin{document}  
\maketitle       
\abstract{
Two critical aspects of radio interferometric imaging analysis are data calibration and deconvolution of the point spread function (PSF) structure.
Both of these are particularly important for high-frequency observations using a VLBI network consisting of a small number of stations, such as those conducted by the Event Horizon Telescope (EHT). 
The Event Horizon Telescope Collaboration (EHTC) has presented images of ring-shaped black holes from observations of \virgoa~($d = 42 \pm 3~\mu \rm as$)~\cite{EHTC1} and the Galactic Center ($d = 51.8 \pm 2.3~\mu \rm as$)\cite{EHTC2022a}. The ring structures seen in the EHTC images are consistent with the estimated shadow diameter of the black hole based on its mass and distance. 
However, these black hole ring sizes are also the same with the typical up and down  spacings (e.g., the intervals between the main beam and nearby 1st-sidelobes) seen in the point spread function (PSF; dirty beam) for each observation. 
These facts suggest that the EHTC ring structures are artifacts derived from the shape of the PSFs rather than the intrinsic structure of the SMBHs in \virgoa and the Galactic Center. 
 The EHTC utilizes novel imaging techniques in addition to the standard CLEAN algorithm. 
The CLEAN method was designed for PSF shape deconvolution in mind, yet in practice, it may not always be able to completely remove the PSF shape. 
In the imaging analysis of data from interferometers with a small number of antennas like the EHT,~it is crucial to assess the PSF shape and compare it with the imaging results. 
The novel imaging methods employed by the EHTC have not yet been fully evaluated for PSF deconvolution performance, and it is highly recommended that their performance in this regard be thoroughly examined.  
It is also important to investigate the data calibration capability, i.e., the ability to separate error noise from the observed data. 
}
%

\section{Principles of VLBI Imaging}
\begin{figure*}[htb!]
\centering  
\includegraphics[width=0.5\textwidth]{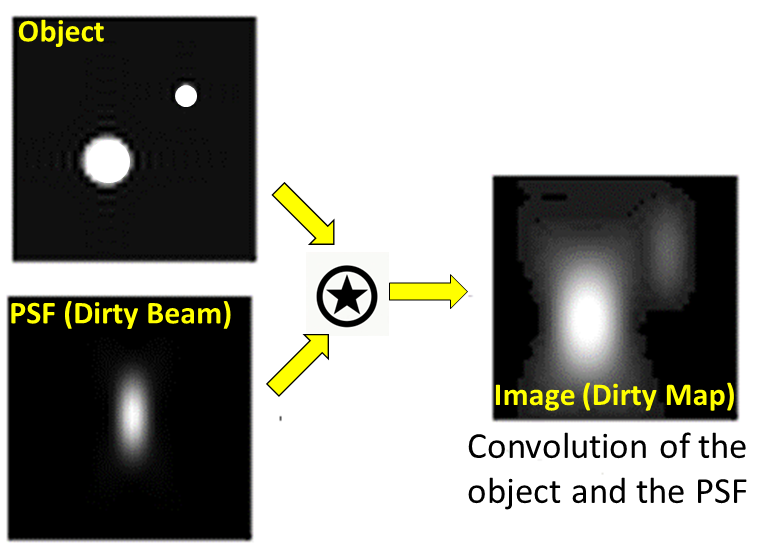}
\caption{The obtained observational image (dirty map) is a convolution of the PSF (dirty beam) of the observing instrument (telescope) with the brightness distribution of the source object.
To obtain the object structure, we need to deconvolve the PSF structure from the dirty map.
}
\label{convolve}
\end{figure*}

Although VLBI imaging appears to use a different approach to data processing than optical/IR telescopes, the basic imaging principle is the same. That is, the resulting image is a convolution of the PSF of the telescope and the structure of the observed object (Figure~\ref{convolve}).
However, in the case of radio interferometry, where the spatial Fourier component cannot be adequately sampled (this is especially true for VLBI), the PSF structure is not sharp, so deconvolution of the PSF structure is necessary to clarify the object image.
In other words, the imaging process in radio interferometry is the process of deconvolving the PSF of the array.
In actual processing, it is difficult to completely remove the influence of the PSF structure, and it often occurs that the influence of the PSF structure may remain.\\
Here, using the EHTC black hole imaging result as a concrete example, we investigate the effects of the PSF structure, in particular their results for the galactic centre. For more information on the independent analysis of the EHT \SGR data, see the paper~\cite{Miyoshi2024b}.

\section{The EHTC Black Hole Ring Images}
We show that
the shapes of the ring image of the EHTC are consistent with the features of its PSF structures.

\subsection{The Main Beam Fits in the EHTC Rings}
\begin{figure*}[htb!]
\centering  
\includegraphics[width=0.38\textwidth]{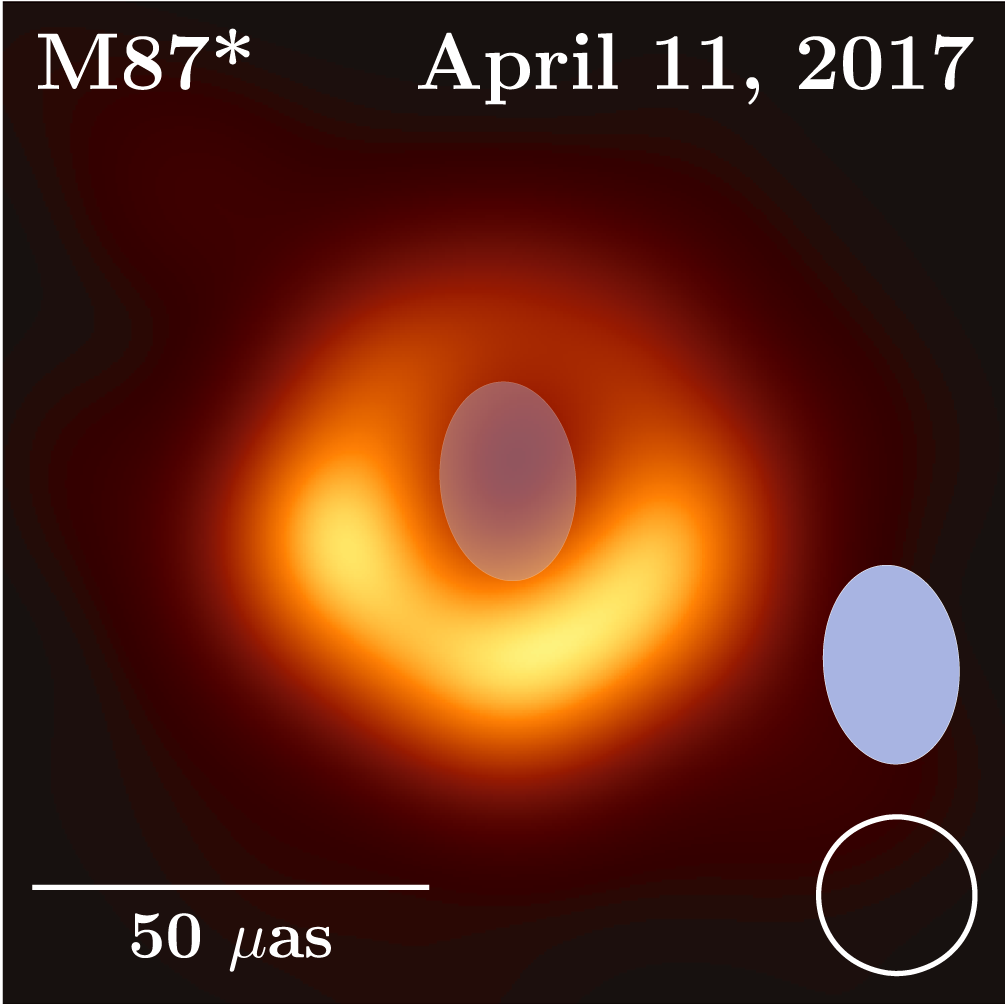}
\includegraphics[width=0.50\textwidth]{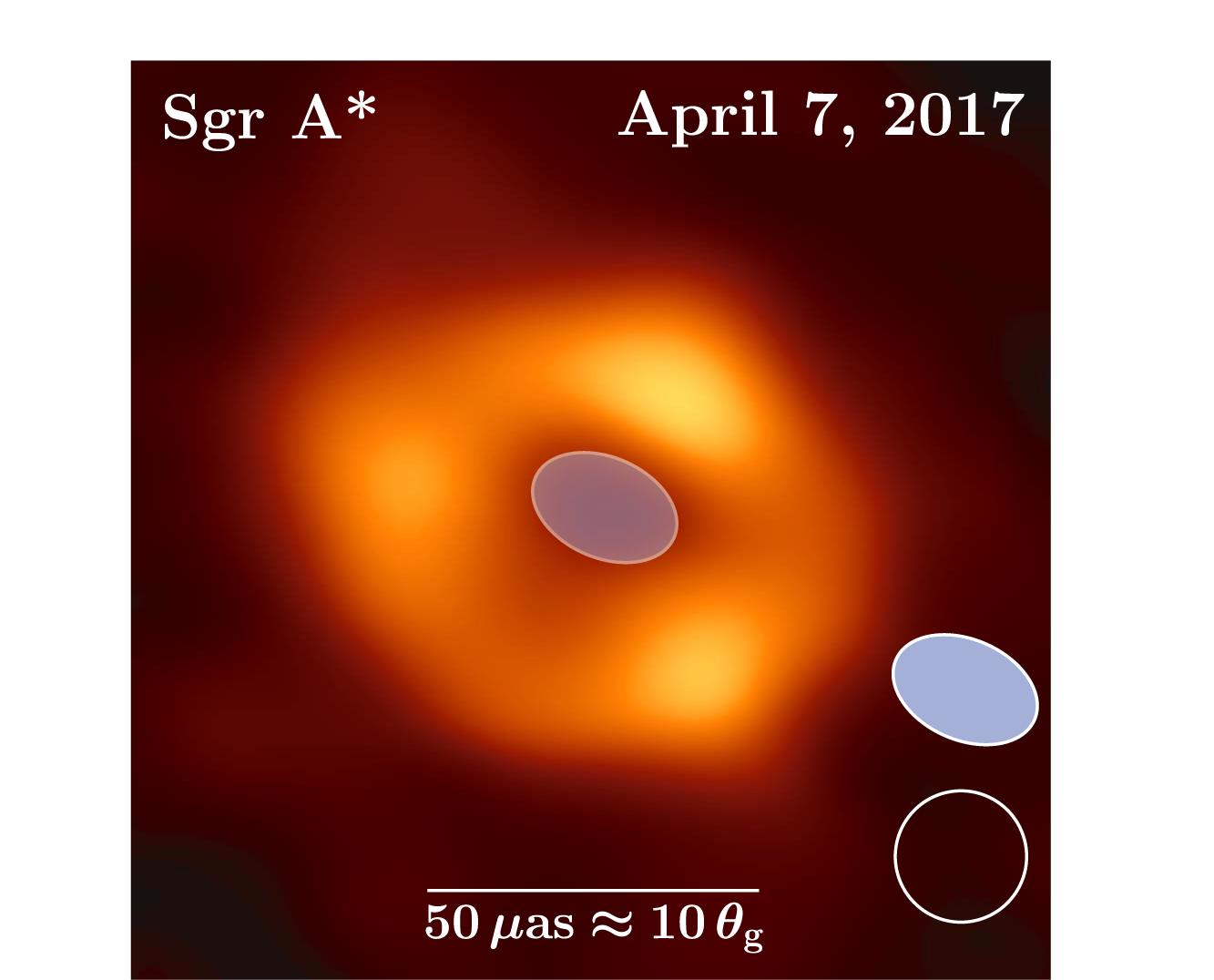}
\caption{Comparison of the default restoring beams and the EHTC images 
The left panel shows the case of \virgoa.
The default restoring beam is an ellipse with $\mathrm{FWHM=25.0\times17.4~\mu \rm as}$ and $\mathrm{PA= 6.0\DEG}$, which shown as a blue ellipse in the panel.
The white circle shows the restoring beam used by EHTC to make their images.
The original image is taken from Figure~3 in~the paper~\cite{EHTC1}.
The left panel shows the case of \SGR.
The default restoring beam is an ellipse with $\mathrm{FWHM=23.0\times15.3~\mu \rm as}$ and $\mathrm{PA= 66.6\DEG}$, which shown as a blue ellipse in the panel.
The white circle shows the restoring beam used by EHTC to make their images.
The original image is taken from Figure~3 in~the paper~\cite{EHTC2022a}
The size and shape of the default restoring beam is nearly the same with the central shadow.}
\label{ring+Nbeam}
\end{figure*}

The influence of PSF is prominently observed in black hole imaging by the EHTC. As shown in Figure~\ref{ring+Nbeam}, the shape of the central hole in the ring image of \SGR reported in the paper~\cite{EHTC2022a} nearly matches the shape of the PSF's main beam. This suggests the possibility of incomplete deconvolution. Note that similar situation has been seen in the \virgoa ~ring image by EHTC~\cite{EHTC1}.

\subsection{Correspondence between PSF Structure and Ring Image}



\begin{figure*}[htb!]
\centering  
\includegraphics[width=\textwidth]{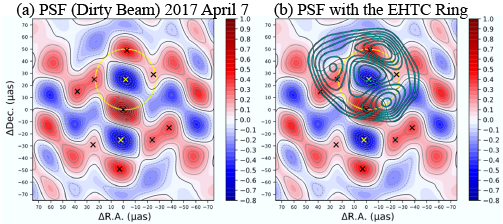}
\caption{
The point spread function (dirty beam) of the EHT array (2017) on the second day of \SGR~observations.
In Panel~(a), the black x-marks represent the peak positions near the center, while the yellow x-marks depict the deepest minimum positions. 
The yellow dotted line indicates the circle with a diameter of $\mathrm{50~\mu \rm as}$ centered on the deepest minimum (north) in the dirty beam.
In Panel~(b), we overlay the EHTC ring image (blue contour lines) on the dirty beam. 
The contour intervals of the EHTC ring image are set at every \textnormal{10~\%} of the peak value.
}\label{ring+PSF}
\end{figure*}

As shown in Figure~\ref{ring+PSF}, examining the PSF structure from the EHT2017 \SGR observational data reveals a bumpy structure around the central main beam, with peaks and dips both spaced at $50~\mu \rm as$ intervals. 

This is consistent with the diameter of the ring shown by the paper~\cite{EHTC2022a}, indicating that the PSF structure significantly affects the imaging results.
The PSFs for both observed sources show very bumpy structures.

The intensity of the first sidelobe is non-negligible with respect to the main beam and a deep dip is present at their midpoint. 
The interval between the bumpy PSF structures is consistent with the diameter of the ring image obtained by EHTC~(Table~\ref{tab:ehtc_measurements}).\\
We have also confirmed that it is possible to create rings that are very similar to EHTC rings, even from simulated data that are not ring structures.
Thus, the EHTC ring image can be considered to originate not from the observed source but from the PSF structure.
\vspace{15mm}
\begin{table*}[]  
\centering  
\resizebox{0.9\textwidth}{!}{  
\begin{tabular}{@{}lcc@{}}
\toprule
 & \virgoa & \SGR \\
\midrule
Predicted Shadow Size & $37.6^{+6.2}_{-3.5}$ or $21.3^{+5}_{-1.7}~\mu\text{as}$& $\sim 50~\mu\text{as}$ \\
\midrule
\multicolumn{3}{@{}l}{EHTC Measurements} \\
\quad $D_{\text{ring}}$ & $42 \pm 3~\mu\text{as}$ & $51.8 \pm 2.3~\mu\text{as}$ \\
\quad $D_{\text{Shadow}}$ & - & $48.7 \pm 7.0~\mu\text{as}$ \\
\midrule
\multicolumn{3}{@{}l}{EHT PSF Structure} \\
\multicolumn{2}{@{}l}{\quad $1^{\text{st}}$ Sidelobe Position from the Main Beam} & \\
\quad & $46~\mu\text{as}$ & $49.09~\mu\text{as}$ \\
\multicolumn{2}{@{}l}{\quad $1^{\text{st}}$ Sidelobe Intensity Relative to the Main Beam} & \\
\quad & $+70~\%$ & $+49~\%$ \\
\multicolumn{2}{@{}l}{\quad Negative Minima at the Midpoint} & \\
\quad & $-60~\%$ & $-78.1~\%$ \\
\midrule
\multicolumn{3}{@{}l}{Restoring Beam Shape} \\
\multicolumn{3}{@{}l}{~~Default} \\
\quad ~~FWHM$_{maj\times min}$& $25.4~\times17.4~\mu\text{as}$ & $23.0\times15.3~\mu\text{as}$ \\
\quad ~~~Position Angle & $6.0\DEG$ & $66\DEG$ \\
~~EHTC Used &         &  \\
\quad ~~~FWHM & $20~\mu \text{as}$         & $20~\mu \text{as}$ \\
\bottomrule
\end{tabular}
}
\caption{
\parbox{0.88\textwidth}{
\vspace{3mm}
Measurements of the EHTC rings and the characteristics of the corresponding PSFs. Predicted shadow sizes, measured ring diameters, and the restoring beam shapes are from EHTC papers. Default beam values are from April 11 for \virgoa and April 7 for \SGR. The values of the PSF structures are from our measurements.
}
}
\label{tab:ehtc_measurements}
\end{table*}

\subsection{EHTC Imaging Simulation Results}
\begin{figure*}[htb!]
\centering  
\includegraphics[width=\textwidth]{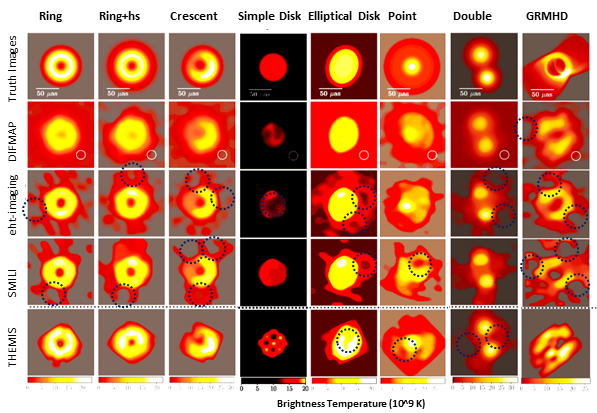}
\caption{Influence of PSF structure found on EHTC imaging simulation results for \SGR.
Influence of the PSF structure found on the EHTC imaging simulation results for \SGR.
This Figure is modified from Fig. 11~(a) showing the EHTC image simulation results in the paper~\cite{EHTC2022b}. 
The image brightness was adjusted for each model image.
The black dotted circles have a diameter of $50~\mu \rm as$. Most of the imaging results show not only the given model image but also something with an interval of about $50~\mu \rm as$. Some of them are ring-shaped with a diameter of $50~\mu \rm as$. Even from the non-ring image models (point, double point and flat brightness disks), the reconstructed images show ring-shaped structures of $50~\mu \rm as$ diameter.
}
\label{PSFin Simulation}
\end{figure*}

The EHTC used several image synthesis algorithms.
As shown in Figure~\ref{PSFin Simulation}, the EHTC image synthesis simulation results from each method show bumpy structures with $50~\mu \rm as$ intervals that are characteristic of the PSF structure. Even from model data that are not ring structures, there are examples of $50~\mu \rm as$ rings or arc-like structures appearing in the resulting images.
This suggests that none of the imaging methods used in EHTC are able to fully deconvolve the effects of PSF structures.

\section{Conclusion}
In VLBI imaging, deconvolution of the PSF structure and calibration of the data are crucial. 
Especially with a limited number of stations, the PSF structure strongly influences the resulting image, highlighting the need for accurate deconvolution. The EHT black hole imaging examples show the significant influence of the PSF on their images.
 Therefore, a detailed analysis of the PSF and its effects must be thoroughly considered in VLBI imaging.

\section{Note}
In principle, we can change the shape of the PSF by changing the weighting of the data points, 
but not much in the case of sparse data sampling; in the \virgoa~and \SGR~data of EHT2017, the separation between the first sidelobe and the main beam is almost unchanged, and the height of the sidelobes is also NOT negligible with respect to the height of the main beam (see, e.g., Figure~32 in~the paper~\cite{Miyoshi2022a}).\\
The EHTC has made data from the first observations (2017) public at the same time as the publication of their paper. We thank the EHTC for this, which allowed us to conduct our independent investigation and analysis. 
We hope that all of subsequent EHT observational data including polarimetric data will be made publicly available as soon as possible.
%
%

%
\end{document}